\documentclass[letterpaper, 11pt]{article}

\usepackage{amsmath, amssymb, amsfonts, amsthm}
\usepackage{bm}
\usepackage{mathtools}
\usepackage{mathrsfs}
\usepackage{geometry}
\usepackage{setspace}
\usepackage[utf8x]{inputenc}
\usepackage{algpseudocode,algorithmicx,algorithm}
\usepackage[T1]{fontenc}
\usepackage{lmodern}
\usepackage{graphicx}
\usepackage{enumerate}
\usepackage[natural, dvipsnames, pdftex]{xcolor}
\usepackage{tikz}
\usepackage{verbatim}
\usepackage{float}
\usepackage{hyperref}
\usepackage{cancel}
\usepackage[all]{xy}
\usepackage{mdframed}
\usepackage[capitalize]{cleveref}
\usepackage{todonotes}
\usepackage{algpseudocode}
\usepackage{comment}
\usepackage{booktabs}
\usepackage{tabularx}
\usepackage{array}
\usepackage{xcolor}
\usepackage{colortbl}

\definecolor{pqhead}{RGB}{232,239,246}
\definecolor{pqrule}{RGB}{157,174,190}

\usepackage[numbers]{natbib}

\usepackage{enumitem}

\usepackage{indentfirst}
\theoremstyle{plain}
\newtheorem{thm}{Theorem}[section]
\newtheorem{lemma}[thm]{Lemma}

\newtheorem{cor}[thm]{Corollary}

\theoremstyle{definition}
\newtheorem{defn}[thm]{Definition}

\theoremstyle{remark}

\newcommand{\ba}{\mathbf{a}}

\newcommand{\F}{\mathbb{F}}

\newcommand{\bx}{\mathbf{x}}
\newcommand{\by}{\mathbf{y}}
\newcommand{\bz}{\mathbf{z}}
\newcommand{\bc}{\mathbf{c}}

\newcommand{\cC}{\mathcal{C}}

\newcommand{\E}{\mathbb{E}}

\newcommand{\ind}{\mathbf{1}}
\newcommand{\RLR}{R_q^{\mathrm{LR}}}
\newcommand{\BLR}{B_q^{\mathrm{LR}}}

\begin{document}

\title{Improving the Trifference Lower Bound via Non-Perfect Hashing
\thanks{
C. Yuan is with School of Computer Science, Shanghai Jiao Tong University. (Email: \href{mailto:chen_yuan@sjtu.edu.cn}{chen\_yuan@sjtu.edu.cn}) R. Zhu is with School of Computer Science, Shanghai Jiao Tong University. (Email: \href{mailto:sjtuzrq7777@sjtu.edu.cn}{sjtuzrq7777@sjtu.edu.cn})
}}

\author{Chen Yuan, Ruiqi Zhu}
\date{}
\maketitle

\begin{abstract}
We strictly improve the K\"orner-Marton lower bound $\frac14\log \frac{9}{5}$ on the asymptotic rate of trifferent codes. We obtain this improvement by constructing non-perfect $(9,3)$-hash codes with rate above $\log(9/5)$ and then applying the Tetra-code construction, which gives $R_3\ge\frac14R_{9,3}$. More generally, for every $q\ge4$, we prove $R_{q,3}>B_{q,3}+\frac1{250000}$. We also extend the method to $(p,2,3)$-list recovery. For every $q\ge4$ and $0\le p<\frac{q-2}{12(q-1)}$, we prove $R_q^{\mathrm{LR}}(p)>B_q^{\mathrm{LR}}(p)$. Here $B_{q,3}$ and $B_q^{\mathrm{LR}}(p)$ denote the respective probabilistic lower bounds. For every fixed $0\le p<1/18$, the rate improvement is bounded below by a positive constant independent of $q\ge4$. For every fixed $0\le p<1/12$, we also obtain a positive rate improvement independent of $q$ for all sufficiently large $q$.
\end{abstract}

\section{Introduction}\label{sec:introduction}
The trifference problem is a classical problem in extremal combinatorics and coding theory. It asks how many ternary vectors can be chosen so that every three distinct vectors have a coordinate with three pairwise distinct entries. Despite its simple formulation, the problem has proved difficult. In particular, the lower bound  on the asymptotic rate established by K\"orner and Marton in 1988~\cite{KornerMarton1988} remained unimproved for almost four decades. In this paper, we give the first strict improvement of this bound.


Let $q,t$ be integers with $2\le t\le q$, and write $[q]=\{1,\ldots,q\}$. A code $\cC\subseteq[q]^N$ is a \emph{$(q,t)$-hash code} if, for every $t$ distinct codewords in $\cC$, some coordinate has $t$ pairwise distinct entries. Such a coordinate is said to \emph{separate} the vectors. When $t=q$, $\cC$ is called a \emph{perfect $q$-hash code} and it is called a \emph{non-perfect $(q,t)$-hash code} when $t<q$. Equivalently, each coordinate gives a function from $\cC$ to $[q]$, and for every $t$ distinct vectors, at least one of these functions takes $t$ distinct values~\cite{FredmanKomlos1984}. Perfect hash families are used in derandomization, particularly in color-coding and algorithms for finding small subgraphs~\cite{AlonYusterZwick1995,NaorSchulmanSrinivasan1995}. Variants of these families are also used in deterministic approximate counting of paths and cycles~\cite{AlonGutner2010}. Perfect hash families and their generalizations have also been studied for applications to cryptography, including frameproof codes and digital fingerprinting~\cite{WalkerColbourn2007,AlonCohenKrivelevichLitsyn2003}.

Hash codes are also closely related to list recovery. In particular, $(q,t)$-hash codes correspond exactly to zero-error list recovery. Here the receiver is given a set $S_i\subseteq[q]$ of at most $t-1$ possible elements at each coordinate. A code is a $(q,t)$-hash code if and only if, for every choice of these sets, at most $t-1$ codewords satisfy $x_i\in S_i$ for all $i\in[N]$. This is precisely $(0,t-1,t)$-list recovery~\cite{ReschYuanZhang2024}.

The rate of a code $\cC\subseteq[q]^N$ is $\log|\cC|/N$, where all logarithms are to base $2$. Let $M_{q,t}(N)$ denote the maximum number of vectors in a $(q,t)$-hash code of length $N$. We define the asymptotic rate of $(q,t)$-hash codes as
\[
R_{q,t}:=
\limsup_{N\to\infty}\frac{\log M_{q,t}(N)}{N}.
\]
The classical probabilistic lower bound~\cite{FredmanKomlos1984,Korner1986,KornerMarton1988} gives
\[
R_{q,t}\ge B_{q,t}:=
-\frac{1}{t-1}
\log\left(1-\frac{q!}{(q-t)!q^t}\right).
\]
For trifferent codes, write $R_3:=R_{3,3}$. K\"orner and Marton~\cite{KornerMarton1988} showed that the ternary Tetra code converts a $(9,3)$-hash code into a trifferent code with four times the length and the same number of vectors. This gives $R_3\ge\frac14R_{9,3}\ge\frac14B_{9,3}=\frac14\log\frac95.$ This bound is stronger than the direct probabilistic bound $B_{3,3}$. We improve it by proving $R_{9,3}>B_{9,3}$.

\subsection{Previous results}

\paragraph{Trifferent codes.}
Han, Sun and Ge~\cite{HanSunGe2026} recently proved that there is a constant $c>0$ such that $M_{3,3}(N)\ge c\sqrt{N}\left(\frac95\right)^{N/4}$ for all sufficiently large $N$. For upper bounds, Bhandari and Khetan~\cite{BhandariKhetan2025} proved $M_{3,3}(N)=O\left(N^{-2/5}(3/2)^N\right).$ They also studied vector trifference~\cite{BhandariKhetanVector2025} and obtained a polynomial factor improvement over a classical upper bound for ordinary trifferent codes. These results improve the known bounds by polynomial factors, but do not change the exponential growth rates. Thus, despite substantial progress on trifferent codes, the asymptotic rate lower bound from 1988 remained unchanged.

Linear trifferent codes have also been studied. Pohoata and Zakharov~\cite{PohoataZakharov2022} proved an upper bound on the number of vectors in linear trifferent codes. Bishnoi et al.~\cite{BishnoiDhaeseleerGijswijtPotukuchi2024} improved this bound and showed that linear trifferent codes can attain the K\"orner-Marton asymptotic rate. Della Fiore and Dalai~\cite{DellaFioreDalai2026} studied upper bounds for linear hash codes with a prescribed number of coordinates at which the entries are pairwise distinct.

\paragraph{Hashing.}

For perfect hashing, K\"orner and Marton~\cite{KornerMarton1988} proved $R_{3,3}>B_{3,3}$ and asked whether $R_{q,q}>B_{q,q}$ holds for every $q\ge3$. Xing and Yuan~\cite{XingYuan2023} proved this for all sufficiently large $q$ and for several smaller values, including $4\le q\le15$. Li and Shi~\cite{LiShi2026} recently proved $R_{q,q}>B_{q,q}$ for every $q\ge3$.

Several works improved the Fredman-Koml\'os upper bound for perfect hashing~\cite{DalaiGuruswamiRadhakrishnan2020,GuruswamiRiazanov2022,CostaDalai2021}. Della Fiore, Costa and Dalai~\cite{DellaFioreCostaDalai2022} also gave upper bounds for general $(q,t)$-hash codes.

Wei, Zhang and Ge~\cite{WeiZhangGe2025}, and later Wei and Ge~\cite{WeiGe2026}, studied separating hash families when the code length is fixed and the alphabet grows. In contrast, we fix $q$ and let the code length tend to infinity.

For non-perfect $(q,3)$-hashing with $q\ge4$, the probabilistic lower bound is $B_{q,3}=\frac12\log\frac{q^2}{3q-2}.$

\paragraph{List recovery.}

Let $q\ge4$. A code $\cC\subseteq[q]^N$ is $(p,2,3)$-list recoverable if, for every choice of $S_1,\ldots,S_N\subseteq[q]$ with $|S_i|\le2$, at most two vectors $\mathbf{x}=(x_1,\ldots,x_N)\in\cC$ satisfy $\bigl|\{i:x_i\notin S_i\}\bigr|\le\lfloor pN\rfloor.$

By Lemma~\ref{lem:list-recovery-separation}, a code is $(p,2,3)$-list recoverable if and only if every three distinct vectors are separated in more than $3\lfloor pN\rfloor$ coordinates. Bishnoi et al.~\cite{BishnoiEtAl2026} also studied ternary codes in which every triple must have several separating coordinates.

Write $\RLR(p)$ for the largest asymptotic rate of $(p,2,3)$-list recoverable codes over $[q]$, and let $a:=(q-1)(q-2)/q^2$. The probabilistic lower bound of Resch, Yuan and Zhang~\cite{ReschYuanZhang2024} gives
\[
\RLR(p)\ge\BLR(p):=
\frac12\left(
3p\log\frac{3p}{a}
+(1-3p)\log\frac{1-3p}{1-a}
\right),
\qquad 0\le3p<a.
\]
The condition $3p<a$ is also necessary for a positive asymptotic rate~\cite{ReschYuanZhang2024}. For every fixed $0\le p<1/3$, Brakensiek et al.~\cite{BrakensiekEtAl2026} proved matching upper and lower bounds showing that $\RLR(p)=\frac{1-3p}{2}\log q+o(\log q)$ as $q\to\infty$.

\subsection{Our results}

Despite the long history of the problem, our proof is short and elementary. We construct an inner code of length $14$ over $[9]$ and show, by a direct probability calculation, that it improves the probabilistic lower bound. Applying the Tetra-code construction of K\"orner and Marton then gives the desired improvement for trifferent codes.

\begin{thm}[Trifference]\label{thm:intro-trifference}
We have $R_3>\frac14\log\frac95.$
\end{thm}
More generally, we obtain the following uniform improvement.
\begin{thm}[Non-perfect hashing]\label{thm:intro-nonperfect}
For every integer $q\ge4$, $R_{q,3}>B_{q,3}+\frac{1}{250000}.$
\end{thm}
We also improve the probabilistic lower bound for list recovery with errors.
\begin{thm}[List recovery]\label{thm:intro-list-recovery}
For every integer $q\ge4$ and every $0\le p<\frac{q-2}{12(q-1)}$, we have $\RLR(p)>\BLR(p).$

Moreover, for every fixed $0\le p<1/18$, there is a constant $c(p)>0$ such that $\RLR(p)\ge\BLR(p)+c(p)$ for all $q\ge4$. For every fixed $0\le p<1/12$, there is a constant $c(p)>0$ such that $\RLR(p)\ge\BLR(p)+c(p)$ for all sufficiently large $q$.
\end{thm}

\subsection{Proof ideas}
We apply the concatenation bound of Xing and Yuan~\cite[Lemma~2.3]{XingYuan2023}, originally proved for $t=q$. The same proof works for $2\le t\le q$ (Lemma~\ref{lem:concatenation}). For an inner code $\cC\subseteq[q]^n$, choose $t$ vectors independently and uniformly from $\cC$, allowing repetitions. Let $F_t(\cC)$ be the probability that no coordinate separates these vectors. Then $R_{q,t}\ge-\frac{\log F_t(\cC)}{n(t-1)}.$ For $\cC=[q]^n$, this gives exactly $B_{q,t}$. Thus, we want to find $\cC$ such that
\[
F_t(\cC)<F_t([q]^n)=\left(1-\frac{q!}{(q-t)!q^t}\right)^n.
\]
Call a tuple \emph{bad} if its entries are not pairwise distinct at any coordinate. We construct the inner code $\cC$ by retaining some vectors from $[q]^n$. A tuple remains only if all its vectors belong to $\cC$. Our goal is to choose $\cC$ so that a larger fraction of bad tuples are removed than of all $t$-tuples in $[q]^n$. This gives $F_t(\cC)<F_t([q]^n)$.

\paragraph{Non-perfect hashing.}
Choose an integer $1\le r<q/2$. Define $s(x)=-1$ for $x\in[r]$ and $s(x)=1$ otherwise, and put $m=\frac1q\sum_{x=1}^q s(x)=1-\frac{2r}{q}\in(0,1).$ We take the inner code
\[
\cC=\left\{\mathbf{x}\in[q]^n:\prod_{i=1}^n s(x_i)=-1\right\}.
\]
Thus, $\cC$ consists of vectors with an odd number of entries in $[r]$.

We now explain why this choice can reduce the failure probability. For three independent uniform vectors $\mathbf{x},\mathbf{y},\mathbf{z}\in[q]^n$, we have $\E\left[\prod_{i=1}^n s(x_i)s(y_i)s(z_i)\right]=m^{3n}.$ Now suppose the triple is bad. At every coordinate, at least two entries are equal. Since $s(x)^2=1$, removing two equal entries leaves one value of $s$. The remaining element is uniform on $[q]$ by symmetry. Since the coordinates remain independent, we obtain $\E\left[\prod_{i=1}^n s(x_i)s(y_i)s(z_i)\,\middle|\,\text{$(\mathbf{x},\mathbf{y},\mathbf{z})$ is bad}\right]=m^n.$

The expression takes only the values $1$ and $-1$. Thus, its probability of being $-1$ is $(1-m^{3n})/2$ for a random triple, but only $(1-m^n)/2$ for a bad triple. Since $0<m<1$, the latter probability is smaller. Every triple from $\cC$ satisfies $\prod_{i=1}^n s(x_i)s(y_i)s(z_i)=-1$. This explains why our choice of $\cC$ may reduce the proportion of bad triples.

The calculation above concerns only $\prod_{i=1}^n s(x_i)s(y_i)s(z_i)$. Having this expression equal to $-1$ does not necessarily mean that all three vectors belong to $\cC$. To compute the probability that all three vectors belong to $\cC$, we use $\mathbf{1}_{\{\mathbf{x}\in\cC\}} =\frac12\left(1-\prod_{i=1}^n s(x_i)\right),$ and the same expression for $\mathbf{y}$ and $\mathbf{z}$. Multiplying these three indicators and expanding gives four types of expectations. Conditioned on the triple being bad, the coordinates are independent, so these expectations can be computed from the corresponding expectations at a single coordinate, see Lemma~\ref{lem:nonperfect-failure}.

We then compare the probability that all three vectors belong to $\cC$, conditioned on the triple being bad, with the same probability for an unrestricted random triple. By choosing suitable $r$ and $n$, we prove that the first probability is smaller, which gives $F_3(\cC)<(1-a)^n$ where $a=\frac{(q-1)(q-2)}{q^2}.$

\paragraph{List recovery.}

For list recovery, we bound the probability that a triple has at most $3\lfloor pN\rfloor$ separating coordinates.

We use the same family of inner codes as for $(q,3)$-hashing, with $r$ and $n$
chosen according to $p$. For three independent uniform codewords $\mathbf{x},\mathbf{y},\mathbf{z}$ from $\cC$, let $J$ count the coordinates where their entries are pairwise distinct, and put $M_{\cC}(w)=\E[w^J]$ where $0\le w<1$. For $0<w<1$, a smaller $J$ gives a larger $w^J$. For $k$ independent inner blocks, the corresponding expectation is $M_{\cC}(w)^k$. Using Markov's inequality and a random sampling and deletion argument similar to that of Xing and Yuan~\cite[Lemma~2.1]{XingYuan2023}, we obtain
\[
\RLR(p)\ge \frac{-\log M_{\cC}(w)+3pn\log w}{2n}.
\]
See Lemma~\ref{lem:list-recovery-concatenation} for the proof.

For $0<p<a/3$, where $a=(q-1)(q-2)/q^2$, we choose $w=\frac{3p(1-a)}{a(1-3p)}.$ For $\cC=[q]^n$, this gives exactly $\BLR(p)$. Thus, to improve this bound, it suffices to prove $M_{\cC}(w)<M_{[q]^n}(w)=(1-a+aw)^n.$ Using the same indicator expansion as in non-perfect hashing, we compute $M_{\cC}(w)$ and prove this inequality, see Lemma~\ref{lem:lr-generating}. For $p=0$, we have $M_{\cC}(0)=F_3(\cC)$, recovering the hashing case.

\subsection{Organization}

Section~\ref{sec:prelims} gives the definitions, the concatenation bounds and the Tetra-code construction. Section~\ref{sec:nonperfect-hashing} constructs non-perfect hash codes and applies the case $q=9$ to improve the trifference lower bound. Section~\ref{sec:list-recovery} extends the method to list recovery.

\paragraph{AI Declaration.}
Apart from limited AI assistance for language editing, no AI assistance was used in obtaining the results of this paper. The authors take responsibility for the correctness of the results and the quality of exposition.

\section{Preliminaries}\label{sec:prelims}
Throughout the paper, $q$ is an integer with $q\ge2$. For a positive integer $m$, write $[m]:=\{1,\ldots,m\}$. All logarithms are to base $2$. For an event $E$, let $\ind_E$ denote its indicator, which is $1$ if $E$ occurs and $0$ otherwise.


\subsection{Hash codes}\label{subsec:hash-codes}

Let $q,t$ be integers with $2\le t\le q$. A $q$-ary code of length $N$ is a nonempty subset $\cC\subseteq[q]^N$. We write $\bx=(x_1,\ldots,x_N)$ for a vector of length $N$. For vectors $\bx_1,\ldots,\bx_t\in[q]^N$, write $(\bx_j)_i$ for the $i$-th coordinate of $\bx_j$. We say that a coordinate $i\in[N]$ \emph{separates} the tuple $(\bx_1,\ldots,\bx_t)$ if $(\bx_1)_i,\ldots,(\bx_t)_i$ are pairwise distinct. Such a coordinate is called a \emph{separating coordinate}. The tuple is \emph{separated} if at least one coordinate separates it. A code $\cC\subseteq[q]^N$ is called a \emph{$(q,t)$-hash code} if every $t$ distinct codewords in $\cC$ are separated. When $t=q$, $\cC$ is called a \emph{perfect $q$-hash code} and it is called a \emph{non-perfect $(q,t)$-hash code} when $t<q$.

Let $M_{q,t}(N)$ denote the maximum number of codewords in a $(q,t)$-hash code of length $N$, and define $R_{q,t}:=\limsup_{N\to\infty} \frac{\log M_{q,t}(N)}{N}.$ For fixed $q,t$, let $a_{q,t}:=\frac{q!}{(q-t)!\,q^t}.$ The well-known probabilistic lower bound~\cite{FredmanKomlos1984,KornerMarton1988,XingYuan2023} states that
\[
R_{q,t}\ge B_{q,t}:=-\frac{1}{t-1}\log(1-a_{q,t}).
\]
Here $B_{q,t}$ is the asymptotic rate obtained by the probabilistic construction of $(q,t)$-hash codes.

For a nonempty code $\cC\subseteq[q]^n$, choose $\mathbf{x}_1,\ldots,\mathbf{x}_t$ independently and uniformly from $\cC$, allowing repetitions. For an integer $K\ge1$, define
\[
F_{t,K}(\cC):=\Pr[\text{fewer than $K$ coordinates separate }(\mathbf{x}_1,\ldots,\mathbf{x}_t)].
\]
For $K=1$, we write $F_t(\cC):=F_{t,1}(\cC) =\Pr[(\mathbf{x}_1,\ldots,\mathbf{x}_t) \text{ is not separated}].$ The following concatenation bound gives a lower bound on $R_{q,t}$ in terms of $F_t(\cC)$.

Xing and Yuan~\cite[Lemma~2.3]{XingYuan2023} proved the following bound for $t=q$. The same argument applies to $2\le t\le q$ and we state the general result below.

\begin{lemma}\label{lem:concatenation}
Fix integers $q,t$ with $2\le t\le q$, and let $\cC\subseteq[q]^n$ be nonempty. If $F_t(\cC)<1$, then $R_{q,t}\ge-\frac{\log F_t(\cC)}{n(t-1)}$.
\end{lemma}

\begin{proof}
Let $m=|\cC|$, and let $\mathcal S(\cC)$ be the collection of all $t$-element subsets of $\cC$ that are separated. Then $F_t(\cC)=1-\frac{t!|\mathcal S(\cC)|}{m^t}.$

Let $\pi$ be any bijection from $\cC$ to $[m]$. Define
\[
\mathcal A:=\bigl\{\{\pi(\bc_1),\ldots,\pi(\bc_t)\}:\{\bc_1,\ldots,\bc_t\}\in\mathcal S(\cC)\bigr\}.
\]
Clearly, $\mathcal A\subseteq\binom{[m]}{t}$ and $|\mathcal A|=|\mathcal S(\cC)|$.

Let $n_1$ be a positive integer. As in~\cite[Lemma~2.1]{XingYuan2023}, choose $\bx_1,\ldots,\bx_t$ independently and uniformly from $[m]^{n_1}$. At each coordinate, their $t$ entries form a set in $\mathcal A$ with probability $t!|\mathcal A|/m^t$. Since the coordinates are independent, the probability that this does not happen at any coordinate is $\left(1-\frac{t!|\mathcal A|}{m^t}\right)^{n_1}.$ Thus, there exists an $m$-ary code $\cC_1\subseteq[m]^{n_1}$ such that every $t$-element subset of $\cC_1$ has a coordinate whose $t$ entries form a set in $\mathcal A$, and its rate satisfies $\frac{\log|\cC_1|}{n_1} \ge-\frac{1}{t-1}\log\left(1-\frac{t!|\mathcal A|}{m^t}\right)-\frac{O(1)}{n_1}.$

Construct a new code $\cC_2$ by replacing each entry of every codeword in $\cC_1$ with its corresponding codeword in $\cC$, namely,
\[
\cC_2:=\left\{\bigl(\pi^{-1}(c_1),\ldots,\pi^{-1}(c_{n_1})\bigr):(c_1,\ldots,c_{n_1})\in\cC_1\right\}.
\]
The code $\cC_2$ has length $nn_1$ and rate $\frac{\log|\cC_2|}{nn_1} \ge-\frac{1}{n(t-1)} \log\left(1-\frac{t!|\mathcal A|}{m^t}\right) -\frac{O(1)}{nn_1}.$

It remains to show that $\cC_2$ is a $(q,t)$-hash code. Choose any $t$ distinct codewords of $\cC_2$, corresponding to $\bc_1,\ldots,\bc_t\in\cC_1$. By the choice of $\cC_1$, there exists $i\in[n_1]$ such that $\{(\bc_1)_i,\ldots,(\bc_t)_i\}\in\mathcal A.$ Hence, $\{\pi^{-1}((\bc_1)_i),\ldots, \pi^{-1}((\bc_t)_i)\}\in\mathcal S(\cC).$ These $t$ codewords are separated, so the corresponding $t$ codewords of $\cC_2$ are also separated. Thus, $\cC_2$ is a $(q,t)$-hash code. Since $|\mathcal A|=|\mathcal S(\cC)|$ and $n$ is fixed, letting $n_1\to\infty$ gives
\[
R_{q,t}\ge-\frac{1}{n(t-1)}\log\left(1-\frac{t!|\mathcal S(\cC)|}{m^t}\right)=-\frac{\log F_t(\cC)}{n(t-1)}.\qedhere
\]
\end{proof}

By Lemma~\ref{lem:concatenation}, to prove $R_{q,t}>B_{q,t}$, it suffices to find a nonempty code $\cC\subseteq[q]^n$ satisfying $F_t(\cC)<(1-a_{q,t})^n.$ The code $\cC$ itself need not be a $(q,t)$-hash code.

\subsection{List recovery}\label{subsec:list-recovery}

We use the following definition from~\cite{ReschYuanZhang2024}.

\begin{defn}\label{def:list-recovery}
Let $0\le p<1$, and let $\ell,L$ be integers with $1\le\ell\le q$ and $L\ge2$. A code $\cC\subseteq[q]^N$ is \emph{$(p,\ell,L)$-list recoverable} if, for every $S_1,\ldots,S_N\subseteq[q]$ with $|S_i|\le\ell$, there are fewer than $L$ codewords $\bx\in\cC$ satisfying $\bigl|\{i\in[N]:x_i\notin S_i\}\bigr|\le pN.$
\end{defn}

For $(p,t-1,t)$-list recovery, we need to count the coordinates that separate a tuple of $t$ vectors. For $\bx_1,\ldots,\bx_t\in[q]^N$, let
\[
d(\bx_1,\ldots,\bx_t):=\bigl|\{i\in[N]:i\text{ separates }(\bx_1,\ldots,\bx_t)\}\bigr|.
\]
For an integer $K\ge1$, the tuple $(\mathbf{x}_1,\ldots,\mathbf{x}_t)$ is \emph{$K$-separated} if $d(\bx_1,\ldots,\bx_t)\ge K$ and it \emph{fails $K$-separation} if $d(\bx_1,\ldots,\bx_t)<K$. For independent uniform samples $\mathbf{x}_1,\ldots,\mathbf{x}_t$ from a nonempty code $\cC$, the failure probability defined above is $F_{t,K}(\cC) =\Pr[d(\mathbf{x}_1,\ldots,\mathbf{x}_t)<K].$

The following lemma shows that $(p,t-1,t)$-list recovery is equivalent to requiring sufficiently many separating coordinates for every $t$ distinct codewords.

\begin{lemma}\label{lem:list-recovery-separation}
Let $2\le t\le q$ and $0\le p<1$. A code $\cC\subseteq[q]^N$ is $(p,t-1,t)$-list recoverable if and only if every tuple of $t$ distinct codewords is $(t\lfloor pN\rfloor+1)$-separated.
\end{lemma}

\begin{proof}
First, suppose that every $t$ distinct codewords are $(t\lfloor pN\rfloor+1)$-separated, but $\cC$ is not $(p,t-1,t)$-list recoverable. Then there exist $t$ distinct codewords $\bx_1,\ldots,\bx_t\in\cC$ and sets $S_1,\ldots,S_N\subseteq[q]$, with $|S_j|\le t-1$ for every $j\in[N]$, such that $\bigl|\{j\in[N]:(\bx_i)_j\notin S_j\}\bigr| \le\lfloor pN\rfloor$ for every $i\in[t]$. At a separating coordinate $j$, the entries $(\bx_1)_j,\ldots,(\bx_t)_j$ are pairwise distinct. Since $|S_j|\le t-1$, at least one of them is not in $S_j$. Therefore, $d(\bx_1,\ldots,\bx_t) \le\sum_{i=1}^t\bigl|\{j\in[N]:(\bx_i)_j\notin S_j\}\bigr| \le t\lfloor pN\rfloor,$ contradicting the assumption.

Conversely, suppose that $\cC$ is $(p,t-1,t)$-list recoverable, but there exist $t$ distinct codewords $\bx_1,\ldots,\bx_t$ with at most $t\lfloor pN\rfloor$ separating coordinates. Partition these coordinates into $t$ groups $J_1,\ldots,J_t$, each containing at most $\lfloor pN\rfloor$ coordinates. For each $j\in J_i$, let $S_j=\{(\bx_k)_j:k\in[t],\,k\ne i\}.$ For each $j\notin\bigcup_{i=1}^tJ_i$, let $S_j=\{(\bx_1)_j,\ldots,(\bx_t)_j\}.$ In both cases, $|S_j|\le t-1$. Moreover, for each $i\in[t]$, we have $(\bx_i)_j\notin S_j$ only if $j\in J_i$. Hence, $\bigl|\{j\in[N]:(\bx_i)_j\notin S_j\}\bigr| \le|J_i|\le\lfloor pN\rfloor.$ Thus, all $t$ codewords agree with the lists in at least $N-\lfloor pN\rfloor$ coordinates, contradicting the assumption that $\cC$ is $(p,t-1,t)$-list recoverable.
\end{proof}

In particular, $(q,t)$-hashing is equivalent to $(0,t-1,t)$-list recovery. For the remainder of this subsection, take $q\ge3$ and $t=3$. Let $M_q^{\mathrm{LR}}(N,p)$ denote the maximum number of codewords in a $(p,2,3)$-list recoverable code of length $N$ over $[q]$, and define $\RLR(p):=\limsup_{N\to\infty}\frac{\log M_q^{\mathrm{LR}}(N,p)}{N}.$ Then $\RLR(0)=R_{q,3}$.

The probabilistic lower bound for list recovery~\cite{ReschYuanZhang2024} gives
\[
\RLR(p)\ge\BLR(p):=\frac12\left(3p\log\frac{3p}{a_{q,3}}+(1-3p)\log\frac{1-3p}{1-a_{q,3}}\right),\qquad 0\le3p<a_{q,3}.
\]
Here $\BLR(p)$ is the asymptotic rate obtained by the probabilistic construction of $(p,2,3)$-list recoverable codes. At $p=0$, we have $\BLR(0)=B_{q,3}$. Moreover, the condition $3p<a_{q,3}$ is necessary for a positive asymptotic rate~\cite{ReschYuanZhang2024}.

The following lemma extends the random sampling and deletion argument of Xing and Yuan~\cite[Lemmas~2.1 and~2.3]{XingYuan2023} to list recovery. Instead of requiring at least one separating coordinate, we now require at least $3\lfloor pN\rfloor+1$ separating coordinates.

Let $\cC\subseteq[q]^n$ be a nonempty code, and choose $\bx,\by,\bz$ independently and uniformly from $\cC$, allowing repetitions. For a real number $w\in[0,1]$, define $J:=d(\bx,\by,\bz)$ and $M_{\cC}(w):=\E[w^J].$ In particular, $M_{\cC}(0)=F_3(\cC)$.

\begin{lemma}\label{lem:list-recovery-concatenation}
Let $\cC\subseteq[q]^n$ be a nonempty code. For every $0\le p<1$ and $0<w<1$,
\[
\RLR(p)\ge\frac{-\log M_{\cC}(w)+3pn\log w}{2n}.
\]
In particular, when $p=0$, we have $\RLR(0)\ge-\frac{\log F_3(\cC)}{2n}.$
\end{lemma}

\begin{proof}
If the claimed lower bound is nonpositive, the result follows from $R_q^{\mathrm{LR}}(p)\ge0$. Otherwise, fix a positive rate $R$ smaller than this bound. For a positive integer $k$, let $M=\lfloor2^{Rnk}\rfloor$.

We choose $M$ vectors independently and uniformly from $\cC^k$, allowing repetitions, where $\cC^k\subseteq[q]^{nk}$ consists of all vectors obtained by concatenating $k$ vectors from $\cC$.

For any three sampled vectors, let $J_i$ denote the number of separating coordinates in their $i$-th block, for $i\in[k]$. Since the blocks are sampled independently and uniformly from $\cC$, the random variables $J_1,\ldots,J_k$ are independent, and each has the same distribution as $J=d(\bx,\by,\bz)$ defined above. Thus, the total number of separating coordinates is $J_1+\cdots+J_k$. By Markov's inequality,
\[
\Pr[J_1+\cdots+J_k\le3\lfloor pnk\rfloor]\le w^{-3\lfloor pnk\rfloor}\E[w^{J_1+\cdots+J_k}]\le \bigl(w^{-3pn}M_{\cC}(w)\bigr)^k.
\]
Thus, the expected number of triples among the $M$ sampled vectors that are separated in at most $3\lfloor pnk\rfloor$ coordinates is at most
\[
\binom{M}{3}\bigl(w^{-3pn}M_{\cC}(w)\bigr)^k\le\frac{M^3}{6}\bigl(w^{-3pn}M_{\cC}(w)\bigr)^k\le\frac{M}{6}\bigl(2^{2Rn}w^{-3pn}M_{\cC}(w)\bigr)^k=o(M),
\]
where the last equality follows from $2^{2Rn}w^{-3pn}M_{\cC}(w)<1,$ by the choice of $R$.

Hence, there exists a choice of the $M$ vectors with at most $o(M)$ such triples. Deleting one vector from each of these triples leaves at least $M-o(M)$ vectors, and every three remaining vectors are separated in at least $3\lfloor pnk\rfloor+1$ coordinates. The remaining vectors are also distinct, since three vectors containing two equal vectors cannot be separated.

By Lemma~\ref{lem:list-recovery-separation}, the remaining vectors form a $(p,2,3)$-list recoverable code of length $nk$. Since the number of remaining codewords is at least $M-o(M)$, its rate approaches $R$ as $k\to\infty$. As $R$ can be arbitrarily close to the claimed lower bound, the result follows.

For $p=0$, the second bound follows directly from Lemma~\ref{lem:concatenation} with $t=3$, since $R_q^{\mathrm{LR}}(0)=R_{q,3}$.
\end{proof}

\subsection{Trifferent codes and the Tetra code}
\label{subsec:trifference}

A \emph{trifferent code} is a perfect $3$-hash code. Its maximum number of codewords at length $N$ is $M_{3,3}(N)$ and its asymptotic rate is $R_3=R_{3,3}$.

We recall the Tetra code of K\"orner and Marton~\cite[Theorem~1]{KornerMarton1988}. Identify $[9]$ with $\F_3^2$ and $[3]$ with $\F_3$. For $\ba=(a_1,a_2)\in\F_3^2$, define $\varphi(\ba):=(a_1,a_2,a_1+a_2,a_1+2a_2).$ The code $\varphi(\F_3^2)$ has nine codewords and is a trifferent code~\cite{KornerMarton1988}. Replacing each entry of a $(9,3)$-hash code of length $N$ by its image under $\varphi$ gives a trifferent code of length $4N$ with the same number of codewords. This gives the following bound.

\begin{lemma}[{\cite{KornerMarton1988,HanSunGe2026}}]
\label{lem:tetra-bound}
For every positive integer $N$, $M_{3,3}(4N)\ge M_{9,3}(N).$ Consequently, $R_3\ge\frac14R_{9,3}.$
\end{lemma}

Together with $B_{9,3}=\log(9/5)$, this gives the classical bound $R_3\ge\frac14\log(9/5)$.


\section{An Improved Trifference Lower Bound via Non-perfect Hashing}
\label{sec:nonperfect-hashing}
We construct non-perfect $(q,3)$-hash codes and use the case $q=9$ to improve the K\"orner-Marton lower bound for trifferent codes. We first compute the failure probability of the inner code. We then prove the trifference result and obtain a rate improvement bounded below by a positive constant for every $q\ge4$.

Fix $q\ge4$, and write $b:=1-a_{q,3}=\frac{3q-2}{q^2}.$ Thus, $b$ is the probability that three independent uniform elements of $[q]$ are not pairwise distinct, and $B_{q,3}=-\frac12\log b$. By Lemma~\ref{lem:concatenation}, it suffices to find a nonempty code $\cC\subseteq[q]^n$ with $F_3(\cC)<b^n$. More precisely, Lemma~\ref{lem:concatenation} gives $R_{q,3}-B_{q,3}\ge-\frac{1}{2n}\log\frac{F_3(\cC)}{b^n}.$

Choose integers $1\le r<q/2$ and $n\ge1$. For $x\in[q]$, define
\[
s(x):=
\begin{cases}
-1, & x\in[r],\\
1, & x\notin[r],
\end{cases}
\qquad m:=\frac1q\sum_{x=1}^q s(x)=1-\frac{2r}{q}.
\]
Then $0<m<1$. We take the inner code
\begin{equation}\label{eq:nonperfect-inner-code}
\cC:=\left\{\bx=(x_1,\ldots,x_n)\in[q]^n:\prod_{i=1}^n s(x_i)=-1\right\}.
\end{equation}
Thus, $\bx\in\cC$ exactly when the number of its entries in $[r]$ is odd.

For a uniformly chosen $\bx\in[q]^n$, we have $\E\left[\prod_{i=1}^n s(x_i)\right]=m^n.$ Since the expression takes only the values $1$ and $-1$, we obtain $|\cC|=\frac{q^n}{2}(1-m^n).$

We next choose three independent uniform vectors from $[q]^n$ and compute $F_3(\cC)$ using conditional probability.

\begin{lemma}\label{lem:nonperfect-failure}
For the code $\cC$ in \eqref{eq:nonperfect-inner-code}, set $u:=\frac{q-2+2qm^2}{3q-2}.$ Then $F_3(\cC)=b^n\,\frac{1-4m^n+3u^n}{(1-m^n)^3}.$
\end{lemma}

\begin{proof}
Choose $\bx,\by,\bz$ independently and uniformly from $[q]^n$, allowing repetitions. By conditional probability,
\[
F_3(\cC)=\Pr[d(\bx,\by,\bz)=0\mid\bx,\by,\bz\in\cC]=b^n\frac{\Pr[\bx,\by,\bz\in\cC\mid d(\bx,\by,\bz)=0]}{\Pr[\bx,\by,\bz\in\cC]}.
\]
Here $\Pr[d(\bx,\by,\bz)=0]=b^n$, since the coordinates are independent. Moreover, $\Pr[\bx,\by,\bz\in\cC] =\left(\frac{|\cC|}{q^n}\right)^3=\frac{(1-m^n)^3}{8}.$ It remains to compute $\Pr[\bx,\by,\bz\in\cC\mid d(\bx,\by,\bz)=0]$.

For each $i\in[n]$, conditioned on $x_i,y_i,z_i$ not being pairwise distinct, each of them is still uniform on $[q]$. Since at least two entries are equal and $s(x)^2=1$ for every $x\in[q]$, the value $s(x_i)s(y_i)s(z_i)$ equals $s$ evaluated at the remaining entry after removing two equal entries. By symmetry under permutations of $[q]$, the remaining entry is uniform on $[q]$. Thus,
\[
\E[s(x_i)]=\E[s(y_i)]=\E[s(z_i)]=\E[s(x_i)s(y_i)s(z_i)]=m,
\]
where all expectations are under this condition.

We now compute the expected value of $s(x_i)s(y_i)$ under the same condition. Before conditioning, the probability that exactly one of $x_i,y_i$ belongs to $[r]$ is $2r(q-r)/q^2$. On this event, $s(x_i)s(y_i)=-1$ and $x_i\ne y_i$. The three entries are not pairwise distinct exactly when $z_i\in\{x_i,y_i\}$, which has conditional probability $2/q$. Thus,
\[
\Pr[s(x_i)s(y_i)=-1\mid x_i,y_i,z_i\text{ are not pairwise distinct}]=\frac{2r(q-r)}{q^2}\cdot\frac{2/q}{b}=\frac{4r(q-r)}{q(3q-2)}.
\]
Since $s(x_i)s(y_i)\in\{-1,1\}$, we obtain
\[
\E[s(x_i)s(y_i)\mid x_i,y_i,z_i\text{ are not pairwise distinct}]=1-\frac{8r(q-r)}{q(3q-2)}=\frac{q-2+2qm^2}{3q-2}=u.
\]
By symmetry, the same holds for $s(x_i)s(z_i)$ and $s(y_i)s(z_i)$.

Conditioned on $d(\bx,\by,\bz)=0$, the triples $(x_i,y_i,z_i)$ are independent over $i\in[n]$. By the calculations above and the definition of $\cC$, we obtain
\begin{align*}
&\Pr[\bx,\by,\bz\in\cC\mid d(\bx,\by,\bz)=0]\\
&\quad=
\frac18\E\Biggl[
\biggl(1-\prod_{i=1}^n s(x_i)\biggr)
\biggl(1-\prod_{i=1}^n s(y_i)\biggr)
\biggl(1-\prod_{i=1}^n s(z_i)\biggr)
\,\Bigm|\,d(\bx,\by,\bz)=0
\Biggr]\\
&\quad=\frac{1-3m^n+3u^n-m^n}{8}=\frac{1-4m^n+3u^n}{8}.
\end{align*}
Substituting these probabilities gives the desired expression for $F_3(\cC)$.
\end{proof}

By Lemma~\ref{lem:nonperfect-failure}, we have $1-\frac{F_3(\cC)}{b^n} = \frac{ m^n\bigl(1-3(u/m)^n+3m^n-m^{2n}\bigr) }{ (1-m^n)^3 }.$ Since $0<m<1$, we have $3m^n-m^{2n}>0$. Thus, $3(u/m)^n<1$ implies $F_3(\cC)<b^n$. To obtain a lower bound on $R_{q,3}-B_{q,3}$, we use the stronger condition $3(u/m)^n<4/5$. In this case, $1-\frac{F_3(\cC)}{b^n}>\frac{m^n}{5}.$ By Lemma~\ref{lem:concatenation} and $-\log x>(1-x)\log e$ for $0<x<1$, we obtain
\begin{equation}\label{eq:nonperfect-gain}
R_{q,3}-B_{q,3}\ge-\frac1{2n}\log\frac{F_3(\cC)}{b^n}>\frac{m^n\log e}{10n}.
\end{equation}

We now take $q=9$ to prove the main trifference result.

\begin{thm}[Trifference]\label{thm:trifference-improvement}
We have
\[
R_3\ge\frac14\log\frac95-\frac1{112}\log\left(\frac{1-4(7/9)^{14}+3(161/225)^{14}}{(1-(7/9)^{14})^3}\right)>\frac14\log\frac95.
\]
In particular, $R_3>0.2120639$.
\end{thm}

\begin{proof}
Take $q=9$, $r=1$ and $n=14$. Then $b=\frac{25}{81},\,m=\frac79,\,u=\frac{161}{225}$ and $\frac um=\frac{23}{25}.$ By Lemma~\ref{lem:nonperfect-failure}, we have $F_3(\cC)=b^{14}\frac{1-4m^{14}+3u^{14}}{(1-m^{14})^3}.$ Since $3(u/m)^{14}=3(23/25)^{14}<1$, the sufficient condition above gives $F_3(\cC)<b^{14}$. Applying Lemma~\ref{lem:concatenation} with $q=9$ and $t=3$, we obtain
\[
\begin{aligned}
R_{9,3}\ge-\frac{\log F_3(\cC)}{28}=\log\frac95-\frac1{28}\log\left(\frac{1-4m^{14}+3u^{14}}{(1-m^{14})^3}\right)>\log\frac95.
\end{aligned}
\]
We now apply Lemma~\ref{lem:tetra-bound}, which gives $R_3\ge\frac14R_{9,3}$. Substituting $m=7/9$ and $u=161/225$ gives the stated lower bound.
\end{proof}

We next use the same inner codes to obtain an improvement that is uniform over all $q\ge4$. We choose $r,n$ so that the lower bound in~\eqref{eq:nonperfect-gain} is at least a positive constant independent of $q$.

\begin{thm}
\label{thm:nonperfect-hashing}
\label{thm:nonperfect-constant}
For every integer $q\ge4$, we have $R_{q,3}>B_{q,3}+\frac1{250000}.$
\end{thm}

\begin{proof}
For $4\le q\le63$, take $r=1$ and $n=2q$. Then $m=1-\frac2q$ and $\frac um=\frac{3q-4}{3q-2}=1-\frac2{3q-2}.$ Using $1-x<e^{-x}$ for $x>0$, we get $3(u/m)^{2q}<3\exp\left(-\frac{4q}{3q-2}\right)<\frac45.$ Since $(1-1/x)^x$ is increasing for $x>1$, taking $x=q/2\ge2$ gives $m^{2q}=\left(1-\frac2q\right)^{2q}\ge\left(\frac12\right)^8=\frac1{256}.$ By \eqref{eq:nonperfect-gain}, we have $R_{q,3}-B_{q,3}>\frac{\log e}{5120q}\ge\frac{\log e}{322560}.$

For $q\ge64$, take $r=\lfloor q/16\rfloor$ and $n=48$. Then $\frac78\le m=1-\frac{2r}{q}<\frac{29}{32}.$ By the formula for $u$ and $m<1$, $\frac um=\frac{q-2+2qm^2}{m(3q-2)}<\frac{1+2m^2}{3m}.$ The last expression is increasing for $m\ge7/8$, so $\frac um<\frac{1+2(29/32)^2}{3(29/32)}<\frac{35}{36}.$ Consequently, $3(u/m)^{48} <3\left(\frac{35}{36}\right)^{48}<\frac45.$ Moreover, $m^{48}\ge\left(\frac78\right)^{48}>\frac1{729},$ since $(8/7)^8<3$. Applying~\eqref{eq:nonperfect-gain}, we have $R_{q,3}-B_{q,3}>\frac{\log e}{10\cdot48\cdot729}=\frac{\log e}{349920}.$

Together, the two cases give $R_{q,3}-B_{q,3} >\frac{\log e}{349920}>\frac1{250000}$ for all $q\ge4$.
\end{proof}

For small values of $q$, we obtain stronger numerical bounds by substituting $r=1$ and suitable values of $n$ into the exact expression in Lemma~\ref{lem:nonperfect-failure} and applying Lemma~\ref{lem:concatenation}. Table~\ref{tab:nonperfect-small-q} lists the resulting bounds for $4\le q\le9$, with $r=1$.

\begin{table}[htbp]
\centering
\caption{
Lower bounds on $R_{q,3}-B_{q,3}$ for small $q$ (values rounded down).
}
\label{tab:nonperfect-small-q}
\small
\setlength{\tabcolsep}{7pt}
\renewcommand{\arraystretch}{1.2}
\begin{tabular}{@{}c c c @{\hspace{0.8cm}} c c c@{}}
\toprule
\multicolumn{3}{c}{$4\le q\le6$}
&
\multicolumn{3}{c}{$7\le q\le9$}\\
\cmidrule(lr){1-3}
\cmidrule(lr){4-6}
$q$ & $n$ & $R_{q,3}-B_{q,3}$
&
$q$ & $n$ & $R_{q,3}-B_{q,3}$\\
\midrule
4 & 5 & $>0.00054519$ & 7 & 11 & $>0.00033399$\\
5 & 7 & $>0.00047705$ & 8 & 12 & $>0.00029039$\\
6 & 9 & $>0.00039672$ & 9 & 14 & $>0.00025892$\\
\bottomrule
\end{tabular}
\end{table}

\section{List Recovery Above the Probabilistic Bound}
\label{sec:list-recovery}
We now consider $(p,2,3)$-list recovery for $q\ge4$. By Lemma~\ref{lem:list-recovery-separation}, a code of length $N$ is $(p,2,3)$-list recoverable if and only if every triple of distinct codewords $\bx,\by,\bz$ satisfies $d(\bx,\by,\bz)>3\lfloor pN\rfloor$. For $p=0$, this is the $(q,3)$-hashing condition studied in Section~\ref{sec:nonperfect-hashing}. We use the same inner codes as in Section~\ref{sec:nonperfect-hashing}. For $p>0$, we need to bound the probability that $d(\bx,\by,\bz)$ is small, rather than only the probability that it equals zero.

Fix $q\ge4$, and write $a:=a_{q,3}=\frac{(q-1)(q-2)}{q^2}$ and $b:=1-a=\frac{3q-2}{q^2}$. We first express the rate improvement in terms of $M_{\cC}(w)$.

For $0<p<a/3$, take $w:=\frac{3pb}{a(1-3p)}$. Then $0<w<1$ and $B_q^{\mathrm{LR}}(p)=\frac{-\log(b+aw)+3p\log w}{2}$. By Lemma~\ref{lem:list-recovery-concatenation}, we obtain
\begin{equation}\label{eq:lr-rate-bound}
R_q^{\mathrm{LR}}(p)-B_q^{\mathrm{LR}}(p) \ge-\frac1{2n}\log\frac{M_{\cC}(w)}{(b+aw)^n}.
\end{equation}

For $p=0$, take $w=0$. Since $M_{\cC}(0)=F_3(\cC)$ and $B_q^{\mathrm{LR}}(0)=B_{q,3}$, the same bound follows from Lemma~\ref{lem:concatenation}. Thus, to prove $R_q^{\mathrm{LR}}(p)>B_q^{\mathrm{LR}}(p)$, it suffices to find a nonempty code $\cC\subseteq[q]^n$ such that $M_{\cC}(w)<(b+aw)^n$. 

We now use the code $\cC$ defined in~\eqref{eq:nonperfect-inner-code}, with $m=1-\frac{2r}{q}$ and $u=\frac{q-2+2qm^2}{3q-2}$. We next compute $M_{\cC}(w)$ for this code.

\begin{lemma}\label{lem:lr-generating}
For the code $\cC$ above, set $u_p:=\left(1-\frac{3p}{a}\right)u+\frac{3p}{a}m^2$ and $v_p:=\left(1-\frac{3p}{a}\right)m+\frac{3p}{a}m^3.$ Then $\frac{M_{\cC}(w)}{(b+aw)^n}=\frac{1-3m^n+3u_p^n-v_p^n}{(1-m^n)^3}.$
\end{lemma}

\begin{proof}
Choose $\bx,\by,\bz$ independently and uniformly from $[q]^n$. We compute $M_{\cC}(w)$ using conditional expectation:
\[
\begin{aligned}
M_{\cC}(w)=\E[w^{d(\bx,\by,\bz)}\mid\bx,\by,\bz\in\cC]=\frac{\E\left[w^{d(\bx,\by,\bz)}
\ind_{\{\bx,\by,\bz\in\cC\}}\right]}
{\Pr[\bx,\by,\bz\in\cC]}.
\end{aligned}
\]
Since the three vectors are independent, we have $\Pr[\bx,\by,\bz\in\cC]=\left(\frac{|\cC|}{q^n}\right)^3=\frac{(1-m^n)^3}{8}.$ Therefore, $M_{\cC}(w)=\frac{8}{(1-m^n)^3}\E\left[w^{d(\bx,\by,\bz)}\ind_{\{\bx,\by,\bz\in\cC\}}\right].$ At coordinate $i$, the contribution to $w^{d(\bx,\by,\bz)}$ is $w$ if $x_i,y_i,z_i$ are pairwise distinct, and $1$ otherwise. Thus,
\[
w^{d(\bx,\by,\bz)}
=\prod_{i=1}^n\left(
 w+(1-w)\ind_{\{x_i=y_i\ \text{or}\ x_i=z_i\ \text{or}\ y_i=z_i\}}
\right).
\]
By the definition of $\cC$,
\[
\ind_{\{\bx,\by,\bz\in\cC\}}
=\frac18\left(1-\prod_{i=1}^n s(x_i)\right)
\left(1-\prod_{i=1}^n s(y_i)\right)
\left(1-\prod_{i=1}^n s(z_i)\right).
\]
Expanding the three factors and using the calculations in the proof of Lemma~\ref{lem:nonperfect-failure}, we obtain
\[
\E\left[w^{d(\bx,\by,\bz)} \ind_{\{\bx,\by,\bz\in\cC\}}\right]=\frac18\Bigl((b+aw)^n-3[m(b+aw)]^n+3[wm^2+(1-w)bu]^n-[wm^3+(1-w)bm]^n\Bigr).
\]
For example, independence across coordinates gives
\[
\E\left[w^{d(\bx,\by,\bz)}\prod_{i=1}^n s(x_i)s(y_i)\right]=\left(
w\E[s(x_i)s(y_i)]
+(1-w)\E\left[
\ind_{\{x_i=y_i\ \text{or}\ x_i=z_i\ \text{or}\ y_i=z_i\}}
s(x_i)s(y_i)
\right]
\right)^n.
\]
By the calculations in the proof of Lemma~\ref{lem:nonperfect-failure}, the two expectations are $m^2$ and $bu$, respectively. Thus, the expression above equals $[wm^2+(1-w)bu]^n.$ The other terms in the expansion are computed in the same way.

By the choice of $w$, we have $\frac{w}{b+aw}=\frac{3p}{a}$ and $\frac{(1-w)b}{b+aw}=1-\frac{3p}{a}.$ Thus, $wm^2+(1-w)bu=(b+aw)u_p$ and $wm^3+(1-w)bm=(b+aw)v_p.$ Substituting these expressions into the expectation above, we obtain
\[
\E\left[w^{d(\bx,\by,\bz)}\ind_{\{\bx,\by,\bz\in\cC\}}\right]=\frac{(b+aw)^n}{8}\left(1-3m^n+3u_p^n-v_p^n\right).
\]
Therefore, $\frac{M_{\cC}(w)}{(b+aw)^n}=\frac{1-3m^n+3u_p^n-v_p^n}{(1-m^n)^3},$ as required.
\end{proof}

Lemma~\ref{lem:lr-generating} gives
\[
1-\frac{M_{\cC}(w)}{(b+aw)^n}=\frac{v_p^n-m^{3n}-3(u_p^n-m^{2n})}{(1-m^n)^3}.
\]
Since $0<m,u<1$ and $0\le3p/a<1$, we have $0<u_p,v_p<1$. Moreover, $v_p-m^3=\left(1-\frac{3p}{a}\right)m(1-m^2)>0.$ If $u_p<v_p$, we can choose $n$ large enough that $3\left(\frac{u_p}{v_p}\right)^n+\left(\frac{m^3}{v_p}\right)^n<1.$ Then $v_p^n>3u_p^n+m^{3n}$, which implies $M_{\cC}(w)<(b+aw)^n$. By~\eqref{eq:lr-rate-bound}, this gives $R_q^{\mathrm{LR}}(p)>B_q^{\mathrm{LR}}(p)$. We now choose $r$ so that $u_p<v_p$.

\begin{thm}\label{thm:lr-main}
Fix $q\ge4$ and $0\le p<\frac{q-2}{12(q-1)}$, and let $h:=\left\lceil\frac{q-2}{q-2-12p(q-1)}\right\rceil$. Then
\[
\RLR(p)>\BLR(p)+\frac{1}{24qh}\left(1-\frac{2+12p}{q}\right)^{2qh}.
\]
\end{thm}

\begin{proof}
Take $r=1$ and $n=2qh$. Write $\eta:=1-\frac{12p(q-1)}{q-2}$. Then $\eta>0$ and $\eta h\ge1$. We have $m=1-\frac2q$, \,$\frac um=1-\frac2{3q-2}$ and $v_p=1-\frac{2+12p}{q}$. By the definitions of $u_p,v_p$, $v_p-u_p=\frac{2(q-2-12p(q-1))}{q(3q-2)}=\eta(m-u)$. Since $v_p\le m$, we obtain $\frac{u_p}{v_p} =1-\frac{v_p-u_p}{v_p}\le1-\frac{2\eta}{3q-2}$. Also, $v_p-m^3=\left(1-\frac{3p}{a}\right)m(1-m^2)$. The assumed range of $p$ gives $\frac{3p}{a}<\frac{q^2}{4(q-1)^2}\le\frac49$. Since $1-m^2=4(q-1)/q^2\ge3/q$ and $v_p\le m$, we have $\frac{m^3}{v_p}\le1-\left(1-\frac{3p}{a}\right)(1-m^2)<1-\frac5{3q}$.

Using $1-x\le e^{-x}$ for $x>0$, together with $n=2qh$ and $\eta h\ge1$, we obtain
\[
3\left(\frac{u_p}{v_p}\right)^n+\left(\frac{m^3}{v_p}\right)^n<3\exp\left(-\frac{4q\eta h}{3q-2}\right)+\exp\left(-\frac{10h}{3}\right)<3e^{-4/3}+e^{-10/3}<\frac56.
\]
Therefore, by Lemma~\ref{lem:lr-generating}, we have
\[
1-\frac{M_{\cC}(w)}{(b+aw)^n}=\frac{v_p^n-m^{3n}-3(u_p^n-m^{2n})}{(1-m^n)^3}>\frac{v_p^n}{6}.
\]
By~\eqref{eq:lr-rate-bound} and $-\log x>1-x$ for $0<x<1$, we have
\[
\RLR(p)-\BLR(p)
\ge-\frac1{2n}\log
\frac{M_{\cC}(w)}{(b+aw)^n}
>\frac{v_p^n}{12n}=\frac1{24qh}
\left(1-\frac{2+12p}{q}\right)^{2qh}.
\]
This proves the theorem.
\end{proof}

For $0\le p<1/18$, we can use the same argument with $(r,n)=(1,2qh)$ for $4\le q\le63$ and $(r,n)=(\lfloor q/16\rfloor,48h)$ for $q\ge64$, where $h$ is defined below. This gives a lower bound independent of $q$.

\begin{cor}\label{cor:lr-uniform}
Fix $0\le p<1/18$, and let $h:=\left\lceil\frac1{1-18p}\right\rceil$. Then, for every integer $q\ge4$, we have $\RLR(p)>\BLR(p)+\frac1{1512h\,3^{8h}}$.
\end{cor}

\begin{proof}
We use the choices of $r,n$ above. Since $(1-18p)h\ge1$, the calculations in Theorem~\ref{thm:nonperfect-hashing} and the formulas for $u_p,v_p$ give $\frac{u_p}{v_p}\le1-\frac{2(1-18p)}{3q-2}$,\, $\frac{m^3}{v_p}<1-\frac5{3q}$,\, $v_p>1-\frac8{3q}$ for $4\le q\le63$. For $q\ge64$, the proof of Theorem~\ref{thm:nonperfect-hashing} gives $\frac78\le m<\frac{29}{32}$,\, $\frac um<\frac{35}{36}$. Also, $a>9/10$ and $3p/a<1/5$. Since $\frac1{36}+m(1-m)\le\frac1{36}+\frac7{64}<\frac a6$, we obtain
\[
\begin{aligned}
\frac{v_p-u_p}{m}
&=\left(1-\frac{3p}{a}\right)
  \left(1-\frac um\right)
  -\frac{3p}{a}m(1-m)\\
&>\frac1{36}
  -\frac{3p}{a}
   \left(\frac1{36}+m(1-m)\right)
 \ge\frac1{36}-\frac p2=\frac{1-18p}{36}.
\end{aligned}
\]
Since $v_p\le m$, this gives $\frac{u_p}{v_p}<1-\frac{1-18p}{36}$. Moreover, $v_p=m\left(1-\frac{3p}{a}(1-m^2)\right)>\frac{4m+m^3}{5}\ge\frac{427}{512}>\frac56$. Since $m^2<(29/32)^2<5/6$, we also have $\frac{m^3}{v_p} <\frac{5m^2}{4+m^2} <\frac{25}{29}<\frac{13}{15}$.

These estimates imply $3\left(\frac{u_p}{v_p}\right)^n+\left(\frac{m^3}{v_p}\right)^n<3e^{-4/3}+e^{-10/3}<\frac56$. For $4\le q\le63$, since $3q/8\ge3/2$ and $(1-1/x)x^x$ is increasing for $x>1$, we have $v_p^{2qh}>\left[\left(1-\frac8{3q}\right)^{3q/8}\right]^{16h/3}\ge3^{-8h}$. For $q\ge64$, we have $v_p>5/6$. Since $(5/6)^6>1/3$, we obtain $v_p^{48h}>\left[(5/6)^6\right]^{8h}>3^{-8h}$. Thus, $v_p^n>3^{-8h}$ in both cases. Since $n\le126h$, Lemma~\ref{lem:lr-generating} gives
\[
1-\frac{M_{\cC}(w)}{(b+aw)^n}=\frac{v_p^n-m^{3n}-3(u_p^n-m^{2n})}{(1-m^n)^3}>\frac{v_p^n}{6}>\frac1{6\cdot3^{8h}}.
\]
By \eqref{eq:lr-rate-bound} and $-\log x>1-x$ for $0<x<1$, we obtain
\[
\RLR(p)-\BLR(p)\ge-\frac1{2n}\log\frac{M_{\cC}(w)}{(b+aw)^n}>\frac1{12n\,3^{8h}}\ge\frac1{1512h\,3^{8h}},
\]
as required.
\end{proof}

For sufficiently large $q$, the same method gives a positive lower bound on $\RLR(p)-\BLR(p)$ for every fixed $p<1/12$.

\begin{cor}\label{cor:lr-large-q}
Fix $0\le p<1/12$, and let $\eta:=1-12p,\,n:=\left\lceil\frac{128}{\eta^2}\right\rceil.$ Then, for every integer $q\ge\lceil64/\eta\rceil$, we have $\RLR(p)>\BLR(p)+\frac1{4n\,2^n}.$
\end{cor}

\begin{proof}
Take $r=\lfloor\eta q/16\rfloor$, with $\eta$ and $n$ as in the statement. Since $q\ge64/\eta$, we have $1-\frac{\eta}{8}\le m<1-\frac{3\eta}{32}$ and $a>1-\frac{3\eta}{64}.$ Since $p=(1-\eta)/12$, we have $\frac{3p}{a}\le\frac{1-\eta}{4-\eta/4}<\frac14.$ Since $m^2<1$, we obtain
\[
\begin{aligned}
\frac{v_p-u_p}{1-m}
&=\left(1-\frac{3p}{a}\right)\frac{m-u}{1-m}
  -\frac{3p}{a}m^2>\frac{1-\eta/4}{3}
  -\frac{3p}{a}
   \left(\frac{1-\eta/4}{3}+1\right)\\
&\ge \frac{1-\eta/4}{3}
  -\frac{1-\eta}{4-\eta/4}
   \frac{4-\eta/4}{3}=\frac{\eta}{4}.
\end{aligned}
\]
Since $1-m>3\eta/32$, it follows that $v_p-u_p>\frac{3\eta^2}{128}.$ Moreover, $v_p>\frac34m\ge\frac{21}{32}>\frac12$ and
\[
v_p-m^3=\left(1-\frac{3p}{a}\right)m(1-m^2)>\frac34\cdot\frac78\cdot\frac{3\eta}{32}
=\frac{63\eta}{1024}
>\frac{3\eta^2}{128},
\]
where the last inequality uses $0<\eta\le1$. Thus,
\[
3\left(\frac{u_p}{v_p}\right)^n
+\left(\frac{m^3}{v_p}\right)^n
\le4\left(1-\frac{3\eta^2}{128}\right)^n
\le4e^{-3}<\frac12.
\]
By Lemma~\ref{lem:lr-generating}, we obtain $1-\frac{M_{\cC}(w)}{(b+aw)^n}>\frac{v_p^n}{2}>\frac1{2^{n+1}}.$ Applying~\eqref{eq:lr-rate-bound} gives $\RLR(p)-\BLR(p)>\frac1{4n2^n},$ as required.
\end{proof}

\bibliographystyle{alpha}
\bibliography{refs}

@article{FredmanKomlos1984,
  author  = {Michael L. Fredman and J{\'a}nos Koml{\'o}s},
  title   = {On the Size of Separating Systems and Families of Perfect Hash Functions},
  journal = {SIAM Journal on Algebraic Discrete Methods},
  volume  = {5},
  number  = {1},
  pages   = {61--68},
  year    = {1984}
}

@article{KornerMarton1988,
  author  = {J{\'a}nos K{\"o}rner and Katalin Marton},
  title   = {New Bounds for Perfect Hashing via Information Theory},
  journal = {European Journal of Combinatorics},
  volume  = {9},
  number  = {6},
  pages   = {523--530},
  year    = {1988}
}

@article{XingYuan2023,
  author  = {Chaoping Xing and Chen Yuan},
  title   = {Beating the Probabilistic Lower Bound on {$q$}-Perfect Hashing},
  journal = {Combinatorica},
  volume  = {43},
  number  = {2},
  pages   = {347--366},
  year    = {2023},
  note    = {Full version: arXiv:1908.08792}
}

@article{ReschYuanZhang2024,
  author  = {Nicolas Resch and Chen Yuan and Yihan Zhang},
  title   = {Zero-Rate Thresholds and New Capacity Bounds for List-Decoding and List-Recovery},
  journal = {IEEE Transactions on Information Theory},
  volume  = {70},
  number  = {9},
  pages   = {6211--6238},
  year    = {2024}
}

@article{Korner1986,
  author  = {J{\'a}nos K{\"o}rner},
  title   = {{Fredman--Koml{\'o}s} Bounds and Information Theory},
  journal = {SIAM Journal on Algebraic Discrete Methods},
  volume  = {7},
  number  = {4},
  pages   = {560--570},
  year    = {1986}
}

@article{AlonYusterZwick1995,
  author  = {Noga Alon and Raphael Yuster and Uri Zwick},
  title   = {Color-Coding},
  journal = {Journal of the ACM},
  volume  = {42},
  number  = {4},
  pages   = {844--856},
  year    = {1995}
}

@article{DalaiGuruswamiRadhakrishnan2020,
  author  = {Marco Dalai and Venkatesan Guruswami and Jaikumar Radhakrishnan},
  title   = {An Improved Bound on the Zero-Error List-Decoding Capacity of the {$4/3$} Channel},
  journal = {IEEE Transactions on Information Theory},
  volume  = {66},
  number  = {2},
  pages   = {749--756},
  year    = {2020}
}

@article{GuruswamiRiazanov2022,
  author  = {Venkatesan Guruswami and Andrii Riazanov},
  title   = {Beating {Fredman--Koml{\'o}s} for Perfect {$k$}-Hashing},
  journal = {Journal of Combinatorial Theory, Series A},
  volume  = {188},
  pages   = {105580},
  year    = {2022}
}

@article{CostaDalai2021,
  author  = {Simone Costa and Marco Dalai},
  title   = {New Bounds for Perfect {$k$}-Hashing},
  journal = {Discrete Applied Mathematics},
  volume  = {289},
  pages   = {374--382},
  year    = {2021}
}

@article{DellaFioreCostaDalai2022,
  author  = {Della Fiore, Stefano and Costa, Simone and Dalai, Marco},
  title   = {Improved Bounds for {$(b,k)$}-Hashing},
  journal = {IEEE Transactions on Information Theory},
  volume  = {68},
  number  = {8},
  pages   = {4983--4997},
  year    = {2022}
}

@article{BhandariKhetan2025,
  author  = {Siddharth Bhandari and Abhishek Khetan},
  title   = {Improved Upper Bound for the Size of a Trifferent Code},
  journal = {Combinatorica},
  volume  = {45},
  year    = {2025},
  note    = {Article 2}
}

@misc{HanSunGe2026,
  author = {Xuejiao Han and Yubo Sun and Gennian Ge},
  title  = {Polynomially Improved Lower Bounds for Trifferent Codes via Locally Sparse {$3$}-Uniform Hypergraphs},
  year   = {2026},
  note   = {arXiv:2607.26376}
}

@article{BishnoiEtAl2026,
  author  = {Anurag Bishnoi and Bart{\l}omiej Kielak and Benedek Kov{\'a}cs and Zolt{\'a}n L{\'o}r{\'a}nt Nagy and G{\'a}bor Somlai and M{\'a}t{\'e} Vizer and Zeyu Zheng},
  title   = {The Generalized Trifference Problem},
  journal = {IEEE Transactions on Information Theory},
  volume  = {72},
  number  = {5},
  pages   = {2907--2914},
  year    = {2026}
}

@misc{BrakensiekEtAl2026,
  author = {Joshua Brakensiek and Yeyuan Chen and Aaron Putterman and Zihan Zhang},
  title  = {Bounds and Limitations on Codes Achieving List Recovery Capacity},
  year   = {2026},
  note   = {arXiv:2607.19576}
}

@article{PohoataZakharov2022,
  author  = {Pohoata, Cosmin and Zakharov, Dmitriy},
  title   = {On the Trifference Problem for Linear Codes},
  journal = {IEEE Transactions on Information Theory},
  volume  = {68},
  number  = {11},
  pages   = {7096--7099},
  year    = {2022}
}

@article{BishnoiDhaeseleerGijswijtPotukuchi2024,
  author  = {Bishnoi, Anurag and D'haeseleer, Jozefien
             and Gijswijt, Dion and Potukuchi, Aditya},
  title   = {Blocking Sets, Minimal Codes and Trifferent Codes},
  journal = {Journal of the London Mathematical Society},
  volume  = {109},
  number  = {6},
  pages   = {e12938},
  year    = {2024}
}

@article{DellaFioreDalai2026,
  author  = {Della Fiore, Stefano and Dalai, Marco},
  title   = {Bounds on {$k$}-Hash Distances and Rates of Linear Codes},
  journal = {IEEE Transactions on Information Theory},
  volume  = {72},
  number  = {2},
  pages   = {865--875},
  year    = {2026}
}

@article{WeiZhangGe2025,
  author  = {Wei, Xin and Zhang, Xiande and Ge, Gennian},
  title   = {Separating Hash Families with Large Universe},
  journal = {Journal of Combinatorial Theory, Series A},
  volume  = {216},
  pages   = {106075},
  year    = {2025}
}

@misc{WeiGe2026,
  author = {Wei, Xin and Ge, Gennian},
  title  = {Separating Hash Families with Large Universe {II}},
  year   = {2026},
  note   = {arXiv:2610.41353}
}

@article{BhandariKhetanVector2025,
  author  = {Bhandari, Siddharth and Khetan, Abhishek},
  title   = {Vector Trifference},
  journal = {Discrete Mathematics},
  volume  = {350},
  number  = {2},
  pages   = {115398},
  year    = {2027}
}

@misc{LiShi2026,
  author = {Li, Shitao and Shi, Minjia},
  title  = {The Probabilistic Bound for Perfect Hashing Is Sharp Only for Binary Alphabets},
  year   = {2026},
  note   = {arXiv:2611.08230}
}

@inproceedings{NaorSchulmanSrinivasan1995,
  author    = {Moni Naor and Leonard J. Schulman
               and Aravind Srinivasan},
  title     = {Splitters and Near-Optimal Derandomization},
  booktitle = {Proceedings of the 36th Annual Symposium
               on Foundations of Computer Science},
  publisher = {IEEE Computer Society Press},
  pages     = {182--191},
  year      = {1995}
}

@article{AlonGutner2010,
  author  = {Noga Alon and Shai Gutner},
  title   = {Balanced Families of Perfect Hash Functions
             and Their Applications},
  journal = {ACM Transactions on Algorithms},
  volume  = {6},
  number  = {3},
  pages   = {54:1--54:12},
  year    = {2010}
}

@article{WalkerColbourn2007,
  author  = {Walker, II, Robert A. and Charles J. Colbourn},
  title   = {Perfect Hash Families: Constructions and Existence},
  journal = {Journal of Mathematical Cryptology},
  volume  = {1},
  number  = {2},
  pages   = {125--150},
  year    = {2007}
}

@article{AlonCohenKrivelevichLitsyn2003,
  author  = {Noga Alon and G{\'e}rard Cohen
             and Michael Krivelevich and Simon Litsyn},
  title   = {Generalized Hashing and Parent-Identifying Codes},
  journal = {Journal of Combinatorial Theory, Series A},
  volume  = {104},
  number  = {1},
  pages   = {207--215},
  year    = {2003}
}


\end{document}